\documentclass[aps,twocolumn,floatfix,a4paper,superscriptaddress]{revtex4-2}
\usepackage{bm,color,graphicx,amsmath,txfonts}

\usepackage[colorlinks, citecolor=blue,linkcolor=blue]{hyperref}
\newcommand{\ic}{{i}}
\newcommand{\e}{{e}}
\begin{document}

\title{Enhancement of magnon-photon-phonon entanglement in a cavity magnomechanics with coherent feedback loop}
\author{Mohamed Amazioug}
\affiliation{LPTHE-Department of Physics, Faculty of sciences, Ibn Zohr University, Agadir, Morocco}
\author{Berihu Teklu}
\affiliation{Department of Applied Mathematics and Sciences, Khalifa University, Abu Dhabi 127788, UAE}
\author{Muhammad Asjad}
\affiliation{Department of Applied Mathematics and Sciences, Khalifa University, Abu Dhabi 127788, UAE}
\begin{abstract}
We propose a scheme to improve magnon-photon-phonon entanglement in cavity magnomechanics using coherent feedback loop. In addition, we prove that the steady state and dynamical state of the system is a genuine tripartite entanglement state. We use the logarithmic negativity as the witness of quantum correlations to quantify the entanglement of all bipartite subsystem, and genuine tripartite entanglement via the nonzero minimum residual contangle, in steady and dynamical regime. We consider the feasible experiment parameters to realize the tripartite entanglement. We show that the entanglement can be significantly improved with coherent feedback using a suitable tuning of the reflective parameter of the beam splitter. The entanglement is robust against the thermal effects. Our proposal scheme to improve the entanglement can be of interest to applications in quantum information.

\end{abstract}

\date{\today}

\maketitle

\section{Introduction}

Cavity optomechanics has attracted significant attention for studying and exploiting the interaction between optical and mechanical degrees of freedom. In optomechanical systems, continuous variable (CV) states (Gaussian states) describe the information encoded in mechanical and optical modes \cite{DVitali2007, asjad18, MAmaziougEPJD2018, asjad15}. In the recent years cavity optomechanical system plays an essential role for studying many interesting phenomena such as quantum entangled states \cite{asjadE13, asjadE16,MAmaziougQIP2020,MAmaziougEPJD2020}. Cooling the mechanical mode to their quantum ground states \cite{JTeufel2011, SMachnes2012, asjadC19, JChan2011, MBhattacharya2007},  photon blockade \cite{MAmaziougPB2022}, generating mechanical quantum superposition states \cite{JQLiao2016, asjad13}, enhancing precision measurements \cite{ZXLiu2017, HXiong20171, HXiong20172}, gravitational-wave detectors \cite{CMCaves1980, AAbramovici1992,VBraginsky2002} and optomechanically induced transpency \cite{eit1,eit2,eit3, eit4} . Quantum state transfer between separate parts is a key tool to achieve quantum information processing protocols and quantum communications \cite{asjad15a, asjad16, asjad22}. Recently, cavity magnomechanics offers a robust platform  where ferrimagnetic cristal (e.g., yttrium iron garnet (YIG) sphere) is coupled with a microwave cavity \cite{DDLachanceQuirion2019, HYYuanArxiv}. We note that the Kittel mode \cite{CKittel1948} in the YIG sphere can realize strong coupling with the microwave photons in a high-quality cavity, leading to cavity polaritons\cite{HHuebl2013,YTabuchi2014,XZhang2014,MGoryachev2014,LBai2015} and the vacuum Rabi splitting. Also, in the cavity magnomechanics, a magnon mode (spin wave) is combined with a vibratory deformation mode of a ferromagnet (or ferrimagnet) by the magnetostrictive force, and a microwave cavity mode by the interaction of magnetic dipoles. The magnetostrictive interaction is a dispersive interaction similar to a radiation pressure for a large ferromagnet, where the frequency of the mechanical mode is very lower than the magnon frequency \cite{XZhang2016,ZYFan2022}. Besides, the first realization of the magnon-photon-phonon interaction \cite{XZhang2016}.

The Entanglement is a significant resource in quantum information processing. This concept was introduced by E. Schrödinger in his replay to the EPR-paradox proposed by A. Einstein et al. \cite{AEinstein1935,ESchrodinger1935}. The entanglement paly a crucial role in various applications in quantum information processing, such as quantum teleportation \cite{CHBennett1993}, superdense coding \cite{CHBennett1992}, telecloning \cite{VScarani2005} and quantum cryptography \cite{AKEkert1991}. The logarithmic negativity \cite{GVidal2002,GAdesso2004} measure the amount of bipartite entanglement systems characterized by continuous variables (CV) of Gaussian states. In this work we consider the coherent feedback loop to improve of entanglement in optomagnomechanical system. This technique is studied theoretically in optomechanical systems \cite{MAmaziougPLA2020, JLi2017, SHuang2019} and recently realized experimentally in optomechanical systems \cite{MRossi2017,JBClark2017,NKralj2017, MRossi2018}.

In this paper, we investigate theoretically the improvement of the entanglement of three bipartites systems and tripartite Gaussian states in an optomagnomechanical system composed of a Fabry-Pérot cavity content inside YIG sphere via coherent feedback as implemented in Fig. 1. A microwave field (not shown) is implemented to improve magnon-phonon coupling. At YIG sphere site, the magnetic field (along x axis) of the cavity mode, the drive magnetic field (in y direction), and bias magnetic field (z direction) are common perpendicular. The coherent feedback technique is presently implemented experimentally in optomechanical systems \cite{MRossi2017,JBClark2017,NKralj2017,MRossi2018}. We employ the logarithmic negativity \cite{GVidal2002,GAdesso2004} to quantify the quantum correlations of three bipartite mode and genuine tripartite entanglement state in stationary-state and in dynamical state. We discuss the evolution of the entanglement of each bipartite Gaussian states and genuine tripartite entanglement state under the effect of the temperature. We demonstrate the role of the feedback technique to make the entanglement robust under the variation of physical parameters characterizing the optomagnechanical system. The first study, demonstrate that the genuine tripartite magnon-phonon-photon entanglement exists in the system, if the magnon mode is in resonance with anti-Stokes (blue sideband) and the cavity mode is in resonance with Stokes (red sideband) \cite{JLi2018}. Magnon squeezing enhanced ground-state cooling in cavity magnomechanics \cite{AsjadMPA}. Entanglement enhancement in cavity magnomechanics by an optical parametric amplifier \cite{BHussain2022}. In this work, we consider the effects of coherent feedback loop on tripartite entanglement.\\

The paper is organized as follows. In Sec. 2, we provide the expression of the Hamiltonian and the corresponding non linear quantum Langevin equations of the opto-magno-mechanical. In Sec. 3, we use linear quantum Lngevin equation and we derive the covariance matrix of the tripartite system. In Sec. 4, we employ the logarithmic negativity to derive the entanglement of three bipartite modes and tripartite mode. The results and discussions are given in Sec. 5. A conclusion closes this paper. 

\begin{figure}[t]\label{fig1}
\hskip-1.0cm\includegraphics[width=1\linewidth]{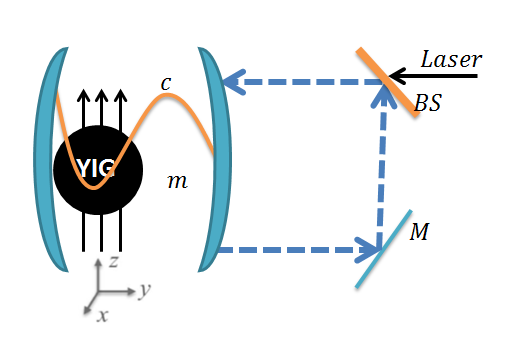} 
\caption{Schematic diagram of a single-mode cavity with feedback loop and a YIG sphere. The magnons are embodied by a collective motion of a large number of spins in a macroscopic ferrimagnet, and the magnon mode is directly driven by a microwave source (not shown) to enhance the magnomechanical coupling. The cavity is also driven by an electromagnetic field with amplitude $\Omega$. The cavity photons and magnons are coupled via magnetic dipole interaction, and the magnons and phonons are coupled via magnetostrictive (radiation pressure-like) interaction.}
\end{figure}

\section{The Model}\label{Model}

The system under study is consists of a cavity magnomechanics drived by single coherent laser source and the microwave. A YIG sphere (a 250-$\mu$m-diameter sphere is used in Ref.~\cite{XZhang2016} is placed inside the cavity, and where the coherent feedback loop is implemented as illustrated in Fig. \ref{fig1}. The magnetic dipole interaction mediates the coupling between magnons and cavity photons. The magnons are coupled with phonons through magnetostrictive interaction. The variable magnetization induced by the magnon excitation inside the YIG sphere results in the deformation of its geometrical structure, which forms vibrational modes (phonons) of the sphere, and vice versa~\cite{CKittel1958}. We consider that the size of the sphere is a lot smaller than the microwave wavelength, so that the influence of the radiation pressure is negligible. The Hamiltonian of the system is given by
\begin{equation} \label{H}
\mathcal{H} = \mathcal{H}_{free} + \mathcal{H}_{md} + \mathcal{H}_{mc} + \mathcal{H}_{dm} + \mathcal{H}_{dc}.
\end{equation}
The first term of $\mathcal{H}$ describes the free magnon and optical modes. Writes as
\begin{equation}
\mathcal{H}_{free} = \hbar\omega_c c^{\dag} c + \hbar\omega_m m^{\dag} m + \frac{\hbar\omega_d}{2} ( q^2 + p^2 ), 
\end{equation}
where $c$ ($c^{\dag}$) and $m$ ($m^{\dag}$) ($[O, O^{\dag}]\,{=}\,1$, $O\,{=}\,c,m$) are the annihilation (creation) operator of the cavity and magnon modes, respectively, $q$ and $p$ ($[q, p]\,{=}\,i$) are the dimensionless position and momentum quadratures of the mechanical mode, and $\omega_c$, $\omega_m$, and $ \omega_d$ are respectively the resonance frequency of the cavity, magnon and mechanical modes. The magnon frequency is determined by the external bias magnetic field $H$ and the gyromagnetic ratio $\gamma$, i.e., $\omega_m=\gamma H$. The second term of Eq. (\ref{H}) is the Hamiltonian describing the interaction between magnon and mechanical modes. It is written by
\begin{equation} 
\mathcal{H}_{md} = \hbar g_{md} m^{\dag} m q. 
\end{equation}
The single-magnon magnomechanical coupling rate $g_{md}$ is small, but the magnomechanical interaction can be improved via driving the magnon mode with a strong microwave field (directly driving the YIG sphere with a microwave source \cite{YPWang2018,YPWang2016}). The third term in Eq. (\ref{H}) gives the interaction between the optical field and the magnon. It reads as
\begin{equation}
\mathcal{H}_{mc} = \hbar g_{mc} (c + c^{\dag}) (m + m^{\dag}).
\end{equation}
The coupling rate $g_{mc}$ between the magnon and microwave can be larger than the dissipation rates $\kappa_c$ and $\kappa_m$ of the cavity and magnon modes respectively, entering into the strong coupling regime, $g_{mc} > \kappa_{c}, \kappa_{m}$~\cite{HHuebl2013,YTabuchi2014,XZhang2014,MGoryachev2014,LBai2015}. In the frame rotating at the drive frequency $\omega_0$ and applying the rotating-wave approximation (RWA), $g_{mc} (c + c^{\dag}) (m + m^{\dag}) \to g_{mc} (c m^{\dag} + c^{\dag} m)$ (valid when $\omega_c, \omega_m \gg g_{mc}, \kappa_{c}, \kappa_{m}$, which is easily satisfied~\cite{XZhang2016}). The five term in the Hamiltonian (1) represent the magnon mode is directly driven by a microwave source (not shown) to enhance the magnomechanical coupling. It is given by 
\begin{equation}
\mathcal{H}_{dm} = i \hbar\mathcal{E} (m^{\dag} e^{-i \omega_0 t}  - m e^{i \omega_0 t} ),
\end{equation}
where $\mathcal{E} =\frac{\sqrt{5}}{4} \gamma \! \sqrt{N} B_0$ is the Rabi frequency ~\cite{JLi2018} describes the coupling strength of the drive magnetic field (with $B_0$ and $\omega_0$ are respectively the amplitude and frequency ) with the magnon mode, where $\gamma/2\pi= 28$ GHz/T, and the total number of spins $N=\rho V$ with $V$ the volume of the sphere and $\rho=4.22 \times 10^{27}$ m$^{-3}$ the spin density of the YIG. The Rabi frequency $\mathcal{E}$ is derived under the hypothesis of the low-lying excitations, $\langle m^{\dag} m \rangle \ll 2Ns$, with $s=\frac{5}{2}$ is the spin number of the ground state Fe$^{3+}$ ion in YIG. The last term in the Hamiltonian (1) characterize the optical field derived of the system which is transmitted by the beam splitter. It is given by
\begin{equation}
\mathcal{H}_{dc} = \hbar\Omega\mu (c^\dagger  e^{i \phi}  -  c  e^{-i \phi} ),
\end{equation}
where where $\phi$ is the phase of electromagnetic field, the quantity $\mu$ and $\tau$ denote the real amplitude transmission parameters of the beam splitter satisfies the equation $\mu^2 + \tau^2 = 1$ ($\mu$ and $\tau$ are real and positive) \cite{MAmaziougPLA2020}. The quantum Langevin equations (QLEs) characterizing the system are given by
\begin{eqnarray} \label{QLEs}
\dot{c}&=& - (i \Delta_{fb} + \kappa_{fb}) c - i g_{mc} m  -i \eta \Omega e^{i \phi}  + \sqrt{2 \kappa_c} c_{fb}^{\rm in}, \nonumber \\
\dot{m}&=& - (i \Delta_m + \kappa_m) m - i g_{mc} c - i g_{md} m q + \mathcal{E} + \sqrt{2 \kappa_m} m^{\rm in},  \nonumber \\
\dot{q}&=& \omega_d p,  \nonumber  \\
\dot{p}&=& - \omega_d q - \gamma_d p - g_{md} m^{\dag}m + \chi, 
\end{eqnarray}
where $\kappa_{fb}=\kappa_c(1-2\tau\cos{\theta})$ and $\Delta_{fb}=\Delta_{c}+2\kappa_c\tau\sin{\theta}$ (with $\Delta_{c}=\omega_{c}-\omega_0$) are respectively the effective cavity decay rate and the detuning with $\alpha$ describes the phase shift generated by the reflectivity of the output field on the mirrors. The operator $C^{in}_{fb}=\tau\e^{\ic\theta}c^{out} +\mu c^{in}$ describes the input optical field induced via the coherent feedback technique. Besides, the output field $c^{out}$ and the cavity field $c$ are related via standard input-output relation $c^{out} = \sqrt{2\kappa_c}c - \mu c^{in}$ \cite{DFWalls1998} (i.e. $C^{in}_{fb}=\tau\sqrt{2\kappa_c}\e^{\ic\theta}c+c^{in}_{fb}$). In addition, the non-zero coherent feedback correlations properties of the input noise operators for the cavity $c^{in}_{fb}$ and $c^{in+}_{fb}$ (where $c^{in}_{fb}=\mu (1-\tau\e^{\ic\theta})c^{in}$) \cite{CWGardiner2000}, are given by 
\begin{eqnarray}
\langle c_{fb}^{\rm in}(t) \, c_{fb}^{\rm in \dag}(t')\rangle &=& \mu^2 (1-\tau\e^{\ic\theta})(1-\tau\e^{-\ic\theta}) [n_c(\omega_c){+}1] \,\delta(t{-}t'), \nonumber \\
\langle c_{fb}^{\rm in \dag}(t) \, c_{fb}^{\rm in}(t')\rangle &=& \mu^2 (1-\tau\e^{\ic\theta})(1-\tau\e^{-\ic\theta}) n_c(\omega_c) \, \delta(t{-}t')
\end{eqnarray}
with $\delta_{m}=\omega_{m}-\omega_0$, $\kappa_m$ is the dissipation rate of the magnon mode, $\gamma_d$ is the mechanical damping rate, and $m^{\rm in}$ and $\xi$ are input noise operators for the magnon and mechanical modes, respectively, which are zero mean and characterized by the following correlation functions~\cite{CWGardiner2000}
\begin{eqnarray}
\langle m^{\rm in}(t) \, m^{\rm in \dag}(t')\rangle &=& [n_m(\omega_m)+1] \, \delta(t{-}t')\nonumber \\
 \langle m^{\rm in \dag}(t) \, m^{\rm in}(t')\rangle &=& n_m(\omega_m)\, \delta(t{-}t')
\end{eqnarray}
and 
\begin{equation}
\langle \chi(t)\chi(t')\,{+}\,\chi(t') \chi(t) \rangle/2 \,\, {\simeq} \,\, \gamma_d [2 n_d(\omega_d) {+}1] \delta(t{-}t')
\end{equation}
The mechanical quality factor ${\cal Q} = \omega_d/\gamma_d \,\, {\gg}\, 1$ is large for a Markovian approximation \cite{Markovian}. Where $n_j(\omega_j){=}\big[ {\rm exp}\big( \frac{\hbar \omega_j}{k_B T} \big) {-}1 \big]^{-1} $ $(j{=}c,m,d)$ are the equilibrium mean thermal photon, magnon, and phonon number, respectively. 
\section{Linearization of quantum Langevin equations}\label{QLEs}
The analytical solution of quantum Langevin equations \ref{QLEs}, can be obtain by using the following linearization scheme $O=O_s +\delta O$ ($O\, {=}\, c,m,q,p$), i.e. we decompose the mode operators as a sum of the steady state average and a fluctuation quantum operator and neglecting second order fluctuation terms when the magnon mode is strongly driven (large amplitude $|m_s| \gg 1$ at the steady state), and the cavity field also has a large amplitude $|a_s| \gg 1$ via the cavity-magnon beamsplitter interaction. This allows us to linearize the dynamics of the system around the steady-state values as
\begin{equation}\label{eq5}
m_s =  \frac{ -ig_{mc}c_s + \mathcal{E} }{ i \Delta_m + \kappa_m }\quad;\quad c_s=-\frac{ig_{mc}m_s+i\mu \Omega \e^{i\phi}}{i\Delta_{fb}+\kappa_{fb}}
\end{equation}
which takes a simpler form 
\begin{equation}\label{avM}
m_s \simeq  \frac{ i  \mathcal{E}  \Delta_{fb} -i\mu\Omega \e^{i\phi}} {g_{ma}^2  -  \Delta_m \Delta_{fb} }\quad;\quad \text{when} \quad | \Delta_m|, |\Delta_{fb}| \gg  \kappa_c, \kappa_m
\end{equation}
where $ \Delta_m = \delta_m + g_{md} q_s$ is the effective magnon-drive detuning including the frequency shift due to the magnomechanical interaction, and $G_{md} = i \sqrt{2} g_{md} m_s$ is the effective magnomechanical coupling rate, where $q_s = - \frac{g_{md}}{\omega_d} m_s^2$.

The linearized QLEs describing the quadrature fluctuations $(\delta Q, \delta P, \delta x, \delta y, \delta q, \delta p)$, with $\delta Q=(\delta c + \delta c^{\dag})/\sqrt{2}$, $\delta P=i(\delta c^{\dag} - \delta c)/\sqrt{2}$, $\delta x=(\delta m + \delta m^{\dag})/\sqrt{2}$, and $\delta y=i(\delta m^{\dag} - \delta m)/\sqrt{2}$, is given by
\begin{equation}
\dot{\Lambda} (t) = \mathcal{F} \Lambda(t) + \nu (t) ,
\end{equation}
where $\Lambda(t)=\big[\delta Q (t), \delta P (t), \delta x (t), \delta y (t), \delta q (t), \delta p (t) \big]^T$, $\nu (t) = \big[ \!\sqrt{2\kappa_c} Q^{\rm in} (t), \sqrt{2\kappa_c} P^{\rm in} (t), \sqrt{2\kappa_m} x^{\rm in} (t), \sqrt{2\kappa_m} y^{\rm in} (t), 0, \chi (t) \big]^T$ is the vector of input noises, and the drift matrix $\mathcal{F}$ can be written as
\begin{equation}\label{AAA}
\mathcal{F} =
\begin{pmatrix}
-\kappa_{fb}  &  \Delta_{fb}  &  0 &  g_{mc}  &  0  &  0  \\
-\Delta_{fb}  & -\kappa_{fb}  & -g_{mc}  & 0  &  0  &  0  \\
0 & g_{mc}  & -\kappa_m  & \tilde{ \Delta}_m & -G_{md} & 0 \\
-g_{mc}  & 0 & -\tilde{ \Delta}_m & -\kappa_m &  0  &  0  \\
0 &  0  &  0  &  0  &  0  &  \omega_d  \\
0 &  0  &  0  &  G_{md}  & -\omega_d & -\gamma_d  \\
\end{pmatrix},
\end{equation}
The drift matrix in Eq.~\eqref{AAA} is provided under the condition $| \Delta_m|, |\Delta_{fb}| \gg  \kappa_c, \kappa_m$. In fact, we will show later that $| \Delta_m|, |\Delta_{fb}| \simeq \omega_d  \gg  \kappa_{fb}, \kappa_m$ [see Fig.~\ref{fig1} (b)] are optimal for the presence of all bipartite entanglements of the system. Note that Eq.~\eqref{eq5} is intrinsically nonlinear since $ \Delta_m$ contains $ |m_s|^2 $. However, for a given value of $ \Delta_m$ (one can always alter $\Delta_m$ by adjusting the bias magnetic field) $m_s$, and thus $G_{md}$, can be achieved straightforwardly.

The time evolution of the quantum fluctuations of the system is a continuous variable (CV) three-mode Gaussian state is completely characterized by a $6\times6$ covariance matrix (CM) $\Gamma$, where $\Gamma_{ij}=\frac{1}{2}\langle \Lambda_i(t) \Lambda_j(t') + \Lambda_j(t') \Lambda_i(t) \rangle$ ($i,j=1,2,...,6$) of the covariance matrix (CM) $\Gamma$ satisfies \cite{DVitali2007, PCParks1993}
\begin{equation}\label{Lyap}
d\Gamma/dt = \mathcal{F} \Gamma + \Gamma \mathcal{F}^T + \mathcal{D},
\end{equation}
where $\mathcal{D}={\rm diag} \big[ \kappa_c \mu^2 (1-\tau)^2 (2n_c+1), \kappa_c\mu^2 (1-\tau)^2 (2n_c+1), \kappa_m (2n_m+1),  \kappa_m (2n_m+1), 0,  \gamma_d (2n_d +1 ) \big]$ is the diffusion matrix, which is defined through $\langle  \nu_i(t) \nu_j(t') + \nu_j(t') \nu_i(t) \rangle/2 = \mathcal{D}_{ij} \delta (t-t')$ and $\Gamma_0 = diag(1,1,1,1,1,1)$ is the CM of the tripartite system at $t=0$.

\section{Quantum correlations}

We adopt the logarithmic negativity to quantify the correlations in the bipartite subsystem in CV system. It is defined by \cite{GVidal2002,GAdesso2004}
\begin{equation} \label{eq:37}
	\mathcal{E}_N = \max[0,-\log(2\xi^-)]
\end{equation}
with $\xi^-$ being the smallest symplectic eigenvalue of partial transposed covariance matrix of two mode Gaussian states 
\begin{equation} \label{eq:38}
\xi^-= \sqrt{\frac{\sigma-\sqrt{\sigma^2-4\det\Gamma}}{2}}   
\end{equation}
The covariance matrix $\Gamma$ associated with the two magnon modes is given by

\begin{equation} \label{eq:Sigmamm}
\Gamma=
\begin{pmatrix}
	\mathcal{A} & \mathcal{C}  \\
    \mathcal{C}^T & \mathcal{B}  
\end{pmatrix}
\end{equation} 
The $2\times 2$ sub-matrices $\mathcal{A}$ and $\mathcal{B}$ in Eq. (\ref{eq:Sigmamm}) describe the autocorrelations of the two magnon modes and $2\times 2$ sub-matrix $\mathcal{C}$ in Eq. (\ref{eq:Sigmamm}) denotes the cross-correlations of the two magnon modes. The symbol $\Gamma$ is written as $\Gamma=\det \mathcal{A}+\det \mathcal{B}-\det \mathcal{C}$. The two subsystems are entangled if $\mathcal{E}_N>0$.

To investigate tripartite entanglement of the system, we use the residual contangle ${\cal R}$~\cite{GAdesso} as quantitative measure, where contangle is a CV analogue of tangle for discrete-variable tripartite entanglement~\cite{WKWootters2000}. A {\it bona fide} quantification of tripartite entanglement is given by the {\it minimum} residual contangle~\cite{GAdesso}
\begin{equation}
{\cal R}_{\rm min} \equiv {\rm min} \Big[ {\cal R}^{c|md}, \, {\cal R}^{m|cd}, \,  {\cal R}^{d|cm}  \Big],
\end{equation}
with ${\cal R}^{i|jk} \equiv C_{i|jk} - C_{i|j} - C_{i|k} \ge 0$ ($i,j,k=c,m,d$) is the residual contangle, with $C_{v|w}$ the contangle of subsystems of $v$ and $w$ ($w$ contains one or two modes), which is a proper entanglement monotone defined as the squared logarithmic negativity \cite{GAdesso}. Besides, a nonzero minimum residual contangle ${\cal R}_{\rm min}\,{>}\,0$ exhibit the existence of {\it genuine} tripartite entanglement in the system. ${\cal R}^{i|jk}>0$ is similar to the Coffman-Kundu-Wootters monogamy inequality~\cite{WKWootters2000} hold for the system of three qubits.

\section{Resultats and Discusion}  

In this section, we will discuss the steady state quantum correlations of two magnon under different effects by considering experimental values reported in \cite{JLi2018}: $\omega_{c}/2\pi= 10$ GHz, $\omega_{d}/2\pi= 10$ MHz , $\gamma_d/2\pi = 100$ Hz, $\kappa_{c}/2\pi = \kappa_{m}/2\pi = 1$ MHz, $g_{mc}/2\pi= G_{md}/2\pi=3.2$ MHz, and and at low temperature $T = 10$ mK. Besides, YIG sphere with a diameter of 0.5 mm was used, which contains more than $10^{17}$ spins. 
\begin{figure}[!htb]
\minipage{0.4\textwidth}
  \includegraphics[width=\linewidth]{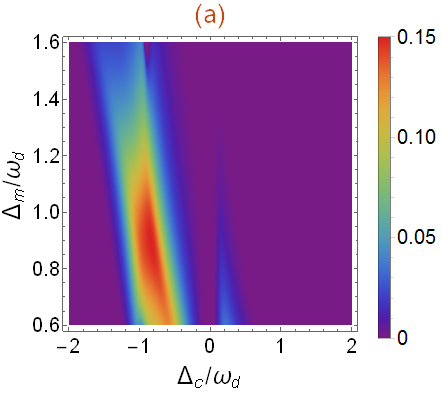}
%  \caption{A really Awesome Image}\label{fig:awesome_image1}
\endminipage\hfill
\minipage{0.4\textwidth}
  \includegraphics[width=\linewidth]{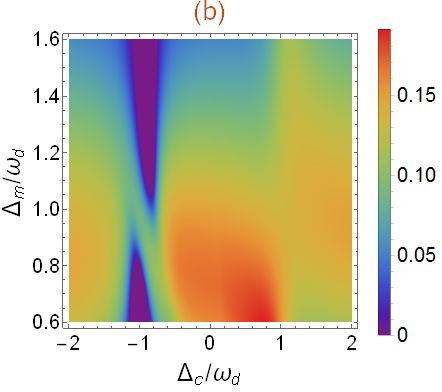}
% \caption{A really Awesome Image}\label{fig:awesome_image2}
\endminipage\hfill
\minipage{0.4\textwidth}
  \includegraphics[width=\linewidth]{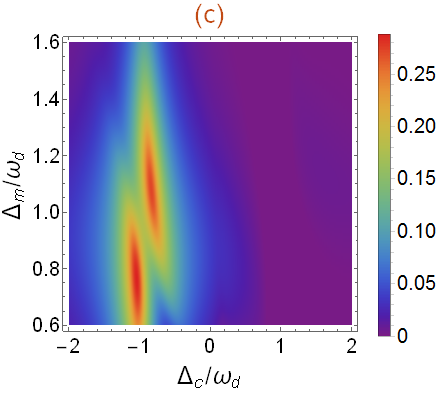}
% \caption{A really Awesome Image}\label{fig:awesome_image2}
\endminipage\hfill
\caption{Plot of bipartite entanglement (a) $Eom$, (b) $EmM$, and (c) $EoM$ versus detunings $\Delta_c$ and $\Delta_m$ with $\tau=0.1$ and $\theta=0$. See text for the other parameters.}
\label{bimodes}
\end{figure}

We plot in Fig. (\ref{bimodes}), the steady state of the three bipartite entanglement $Eom$ (between the cavity and magnon mode), $EmM$ (between the magnon and mechanical mode) and $EoM$ (between the cavity and mechanical mode) as a function of the detunings $\Delta_c$ and $\Delta_m$ in the presence of coherent feedback loop. We remark, the maximum value of entanglement of the three bipartite is enhances by coherent feedback loop in comparison with results in Ref. \cite{JLi2018}. This can explain by the re-injection of the photon in the cavity which improves the coupling between different bipartite modes. We observe, when $\Delta_c = - \Delta_d$ the entanglement $Eom$ and $EoM$ is maximum while the entanglement $EmM$ is 0.10. 

\begin{figure}[!htb]
\minipage{0.4\textwidth}
  \includegraphics[width=\linewidth]{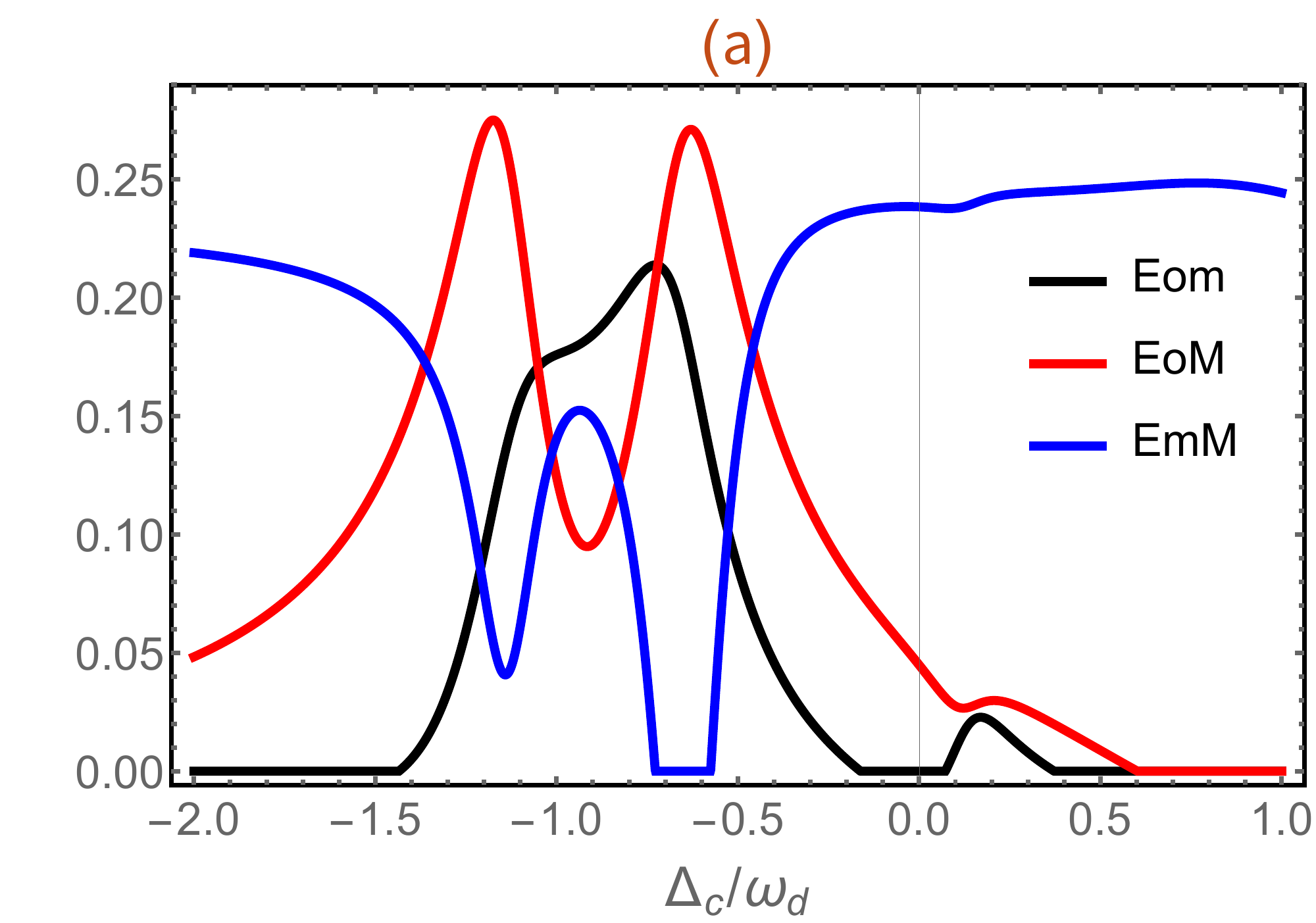}
%  \caption{A really Awesome Image}\label{fig:awesome_image1}
\endminipage\hfill
\minipage{0.4\textwidth}
  \includegraphics[width=\linewidth]{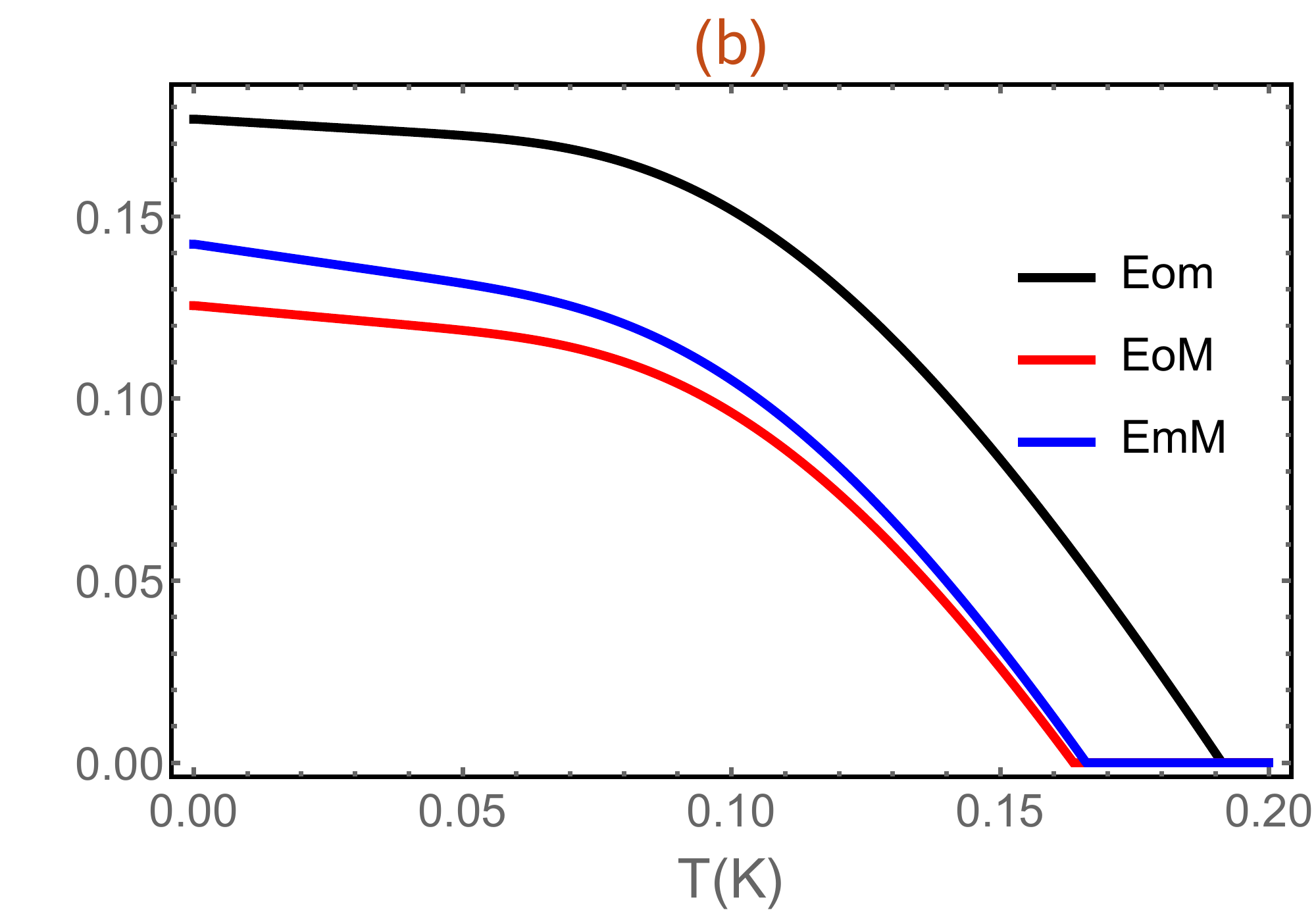}
% \caption{A really Awesome Image}\label{fig:awesome_image2}
\endminipage\hfill
\minipage{0.4\textwidth}
  \includegraphics[width=\linewidth]{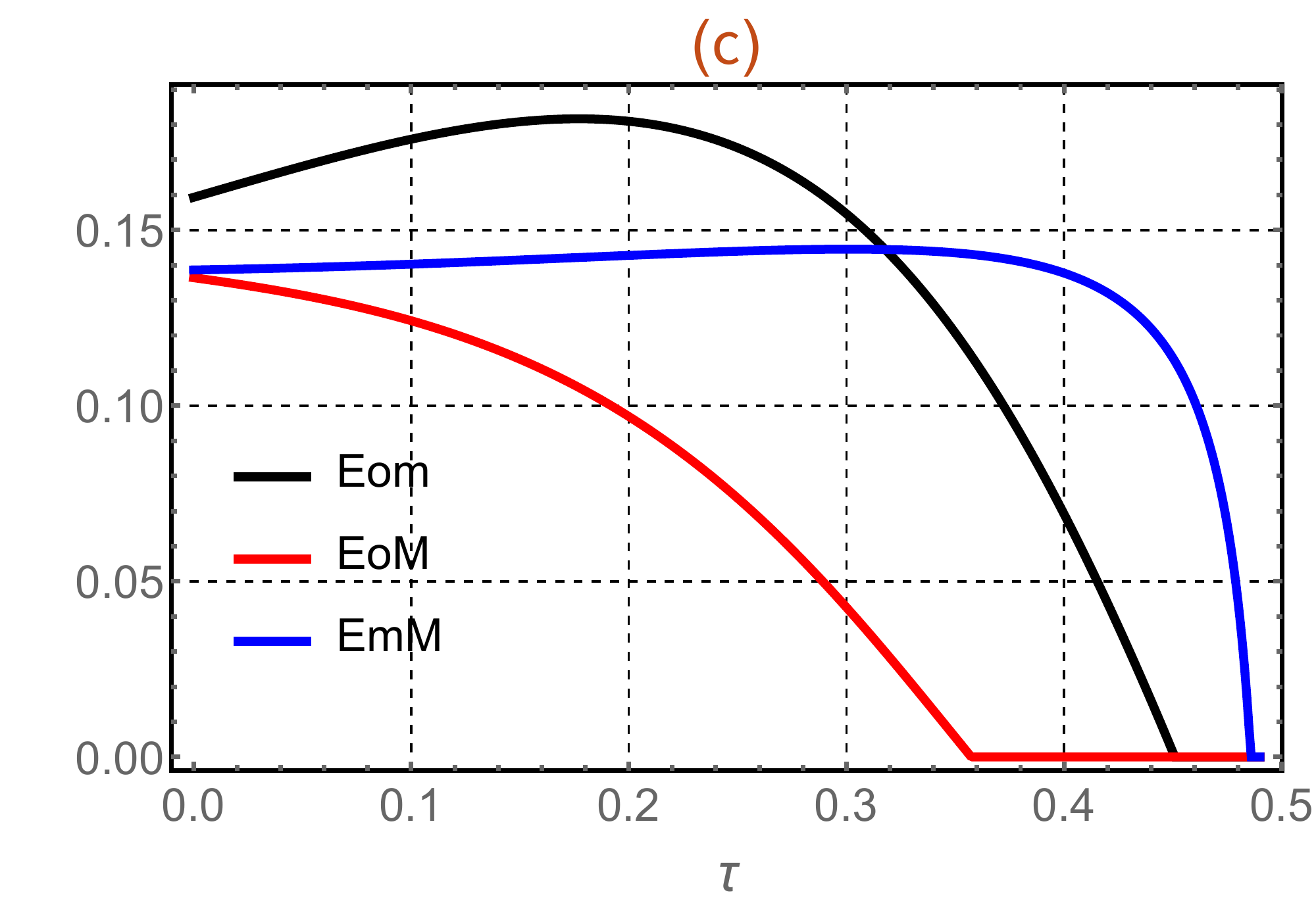}
% \caption{A really Awesome Image}\label{fig:awesome_image2}
\endminipage\hfill
\caption{(a) Plot of $Eom$, $EoM$ and $EmM$ as a function of $\Delta_c/\omega_d$, temperature (see the figure (b)) and reflectivity $\tau$ (see the figure (c)). We take $G_{md}/2\pi = 4.8$ MHz and $\Delta_m = 0.9 \omega_d$. The other parameters are as in Fig. 2;$\tau=0.1$ (a-b) $\Delta_c = - \omega_d$ (a-c). See text for the other parameters.}
\label{bimodes}
\end{figure}

In Fig. \ref{bimodes}, we present the steady state of the three bipartite entanglements $Eom$, $EoM$ and $EmM$ versus different parameters. The three bipartite entanglements are all not vanishing in different regions of $\Delta_c/\omega_d$ in Fig. \ref{bimodes}(a). This means the presence of tripartite entanglement between photon-magnon-phonon. We remark that the three bipartite entanglement are robust against temperature and survive up to about 200 mK (see Fig.\ref{bimodes} (b)) as also discussed in Ref. \cite{JLi2018}. We can explain the diminishing of all bipartite entanglement by the thermal effects induced by decoherence phenomenon \cite{decoherence}. Besides, the two bipartite entanglement $Eom$ and $EoM$ are enhanced with increasing values of the reflectity parameters $\tau$ (i.e. decay rate $\kappa_{fb}$ reduces) and begin to decrease quickly after reach its maximum entanglement value, i.e. one can say that the coherent feedback enhances the bipartite entanglement as in Fig. \ref{bimodes}(c). Moreover, the entanglement between magnon and phonon is decreases quickly with increasing $\tau$. This can be explained by the decoherence effects produce with re-injection of photons in the cavity, because increasing the photon number is responsible of more thermal effects which induce the degradation of the quantum correlations between the two modes as also discussed in Ref. \cite{MAmaziougPLA2020}.

\begin{figure}[!htb]
%\minipage{0.5\textwidth}
  \includegraphics[width=0.4\textwidth]{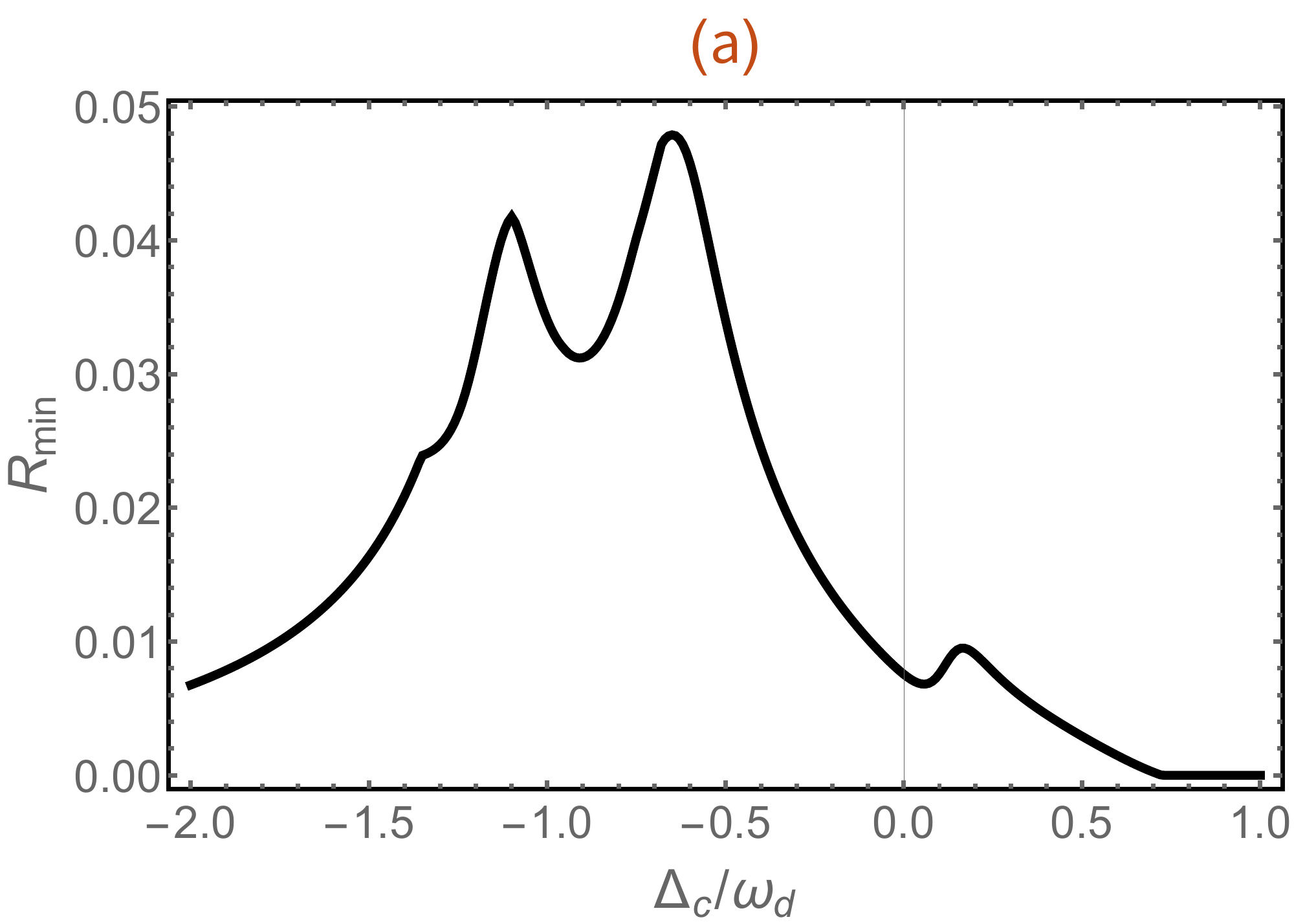}
%  \caption{A really Awesome Image}\label{fig:awesome_image1}
%\endminipage\hfill
%\minipage{0.5\textwidth}
  \includegraphics[width=0.4\textwidth]{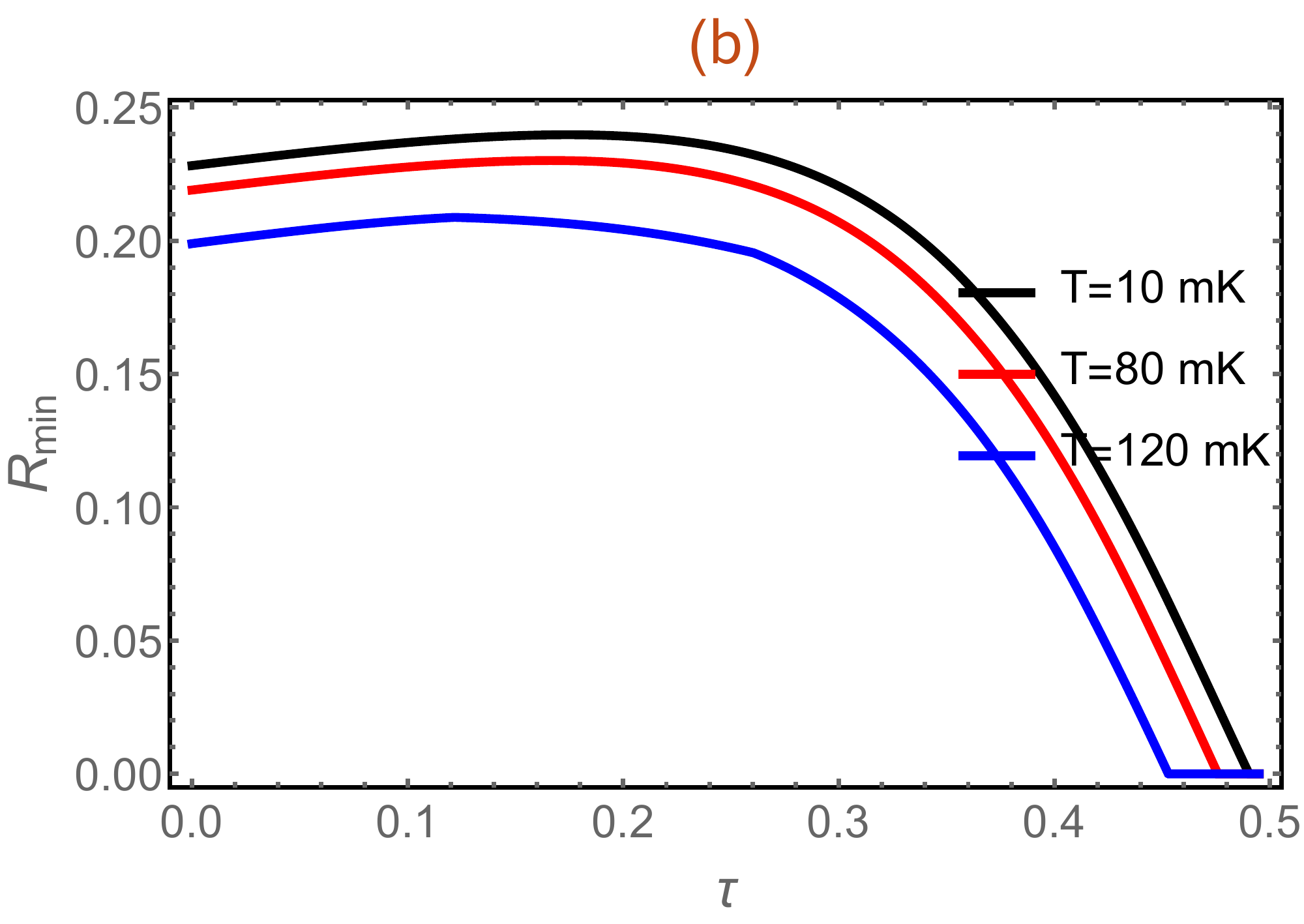}
%  \caption{A really Awesome Image}\label{fig:awesome_image2}
%\endminipage\hfill
\caption{Plot of tripartite entanglement in terms of the minimum residual contangle Rmin versus $\Delta_c$. We take $G_{md}/2\pi = 4.8$ MHz, $\Delta_m = 0.9 \omega_b$, $\Delta_c = - \omega_b$ and $\theta=0$. The other parameters are as in Fig. (a); $\tau=0.2$. See text for the other parameters}
\label{trimodes}
\end{figure}
%.5\textwidth
We plot in Fig. \ref{trimodes}(a) we plot in steady state the minimum of the residual contangle $R_{min}$ versus of detuning $\Delta_c/\omega_d$ with $G_{md}/2\pi = 4.8$ MHz as in Ref. \cite{JLi2018}, for a fixed value of all other parameters. We notice that the system is a genuinely tripartite entangled state as shown by the nonzero minimum residual contangle $R_{min}$ in Fig.\ref{trimodes} (b). Also we plot the evolution of the minimum of the residual contangle versus the reflectivity $\tau$ for different values of the temperature $T$ as implemented in Fig. \ref{trimodes}(b). Firstly, we remark the enhancement of tripartite entanglement with increasing the parameter $\tau$, i.e., the coherent feedback loop enhances tripartite entanglement. This tripartite entanglement is decreases quickly after reaching its maximum value for a specific value of $\tau$. Besides, the $R_{min}$ decreases with increasing the temperature (decoherence phenomenon), i.e. a higher temperature reduces the amount of the tripartite entanglement. Also the region in which tripartite entanglement exists increases with decreasing temperature as shown in Fig. \ref{trimodes}(b).

\begin{figure}[!htb]
\minipage{0.5\textwidth}
  \includegraphics[width=\linewidth]{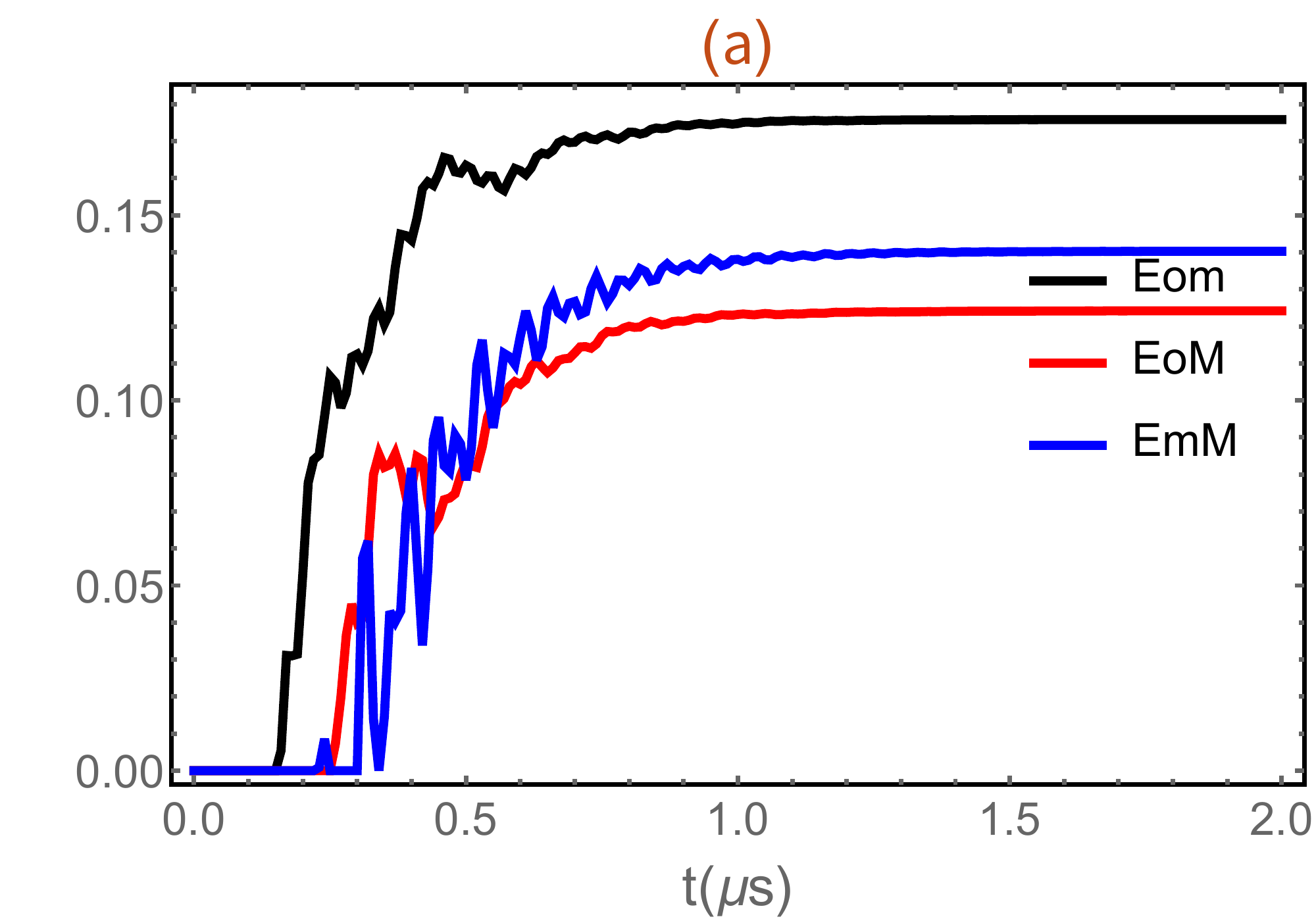}
 % \caption{A really Awesome Image}\label{fig:awesome_image2}
\endminipage\hfill
\minipage{0.5\textwidth}
  \includegraphics[width=\linewidth]{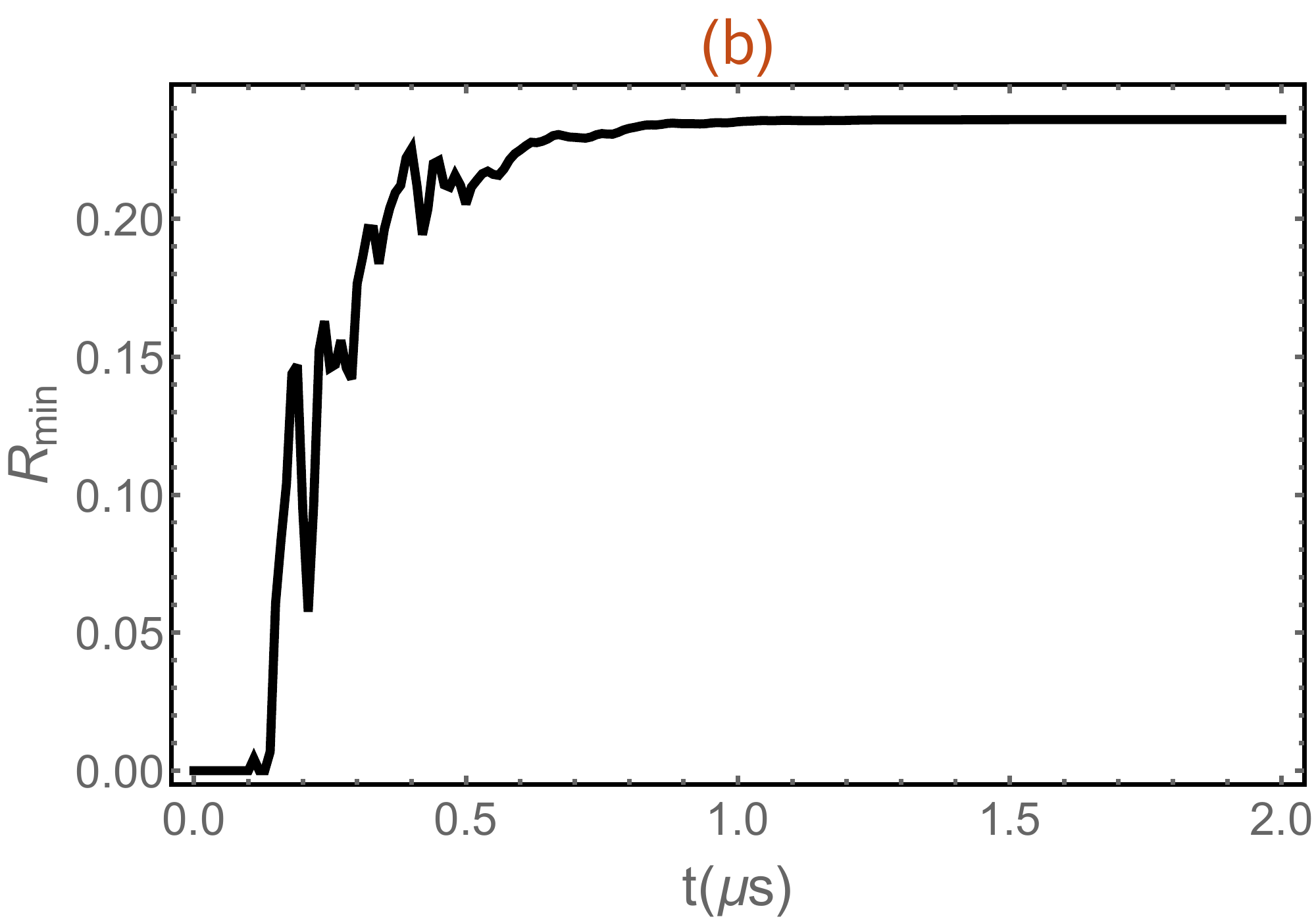}
 % \caption{A really Awesome Image}\label{fig:awesome_image2}
\endminipage\hfill
\caption{Plot of time evolution of the all bipartite entanglement $Eom$, $EoM$ and $EmM$ with $G_{mb}/2\pi = 4.8$ MHz, $\tau=0.1$, $\Delta_c =-\omega_d$ and $\theta=0$. See text for the other parameters.}
\label{dynamics}
\end{figure}

We plot in Fig. \ref{dynamics}(a) time-evolution of the three bipartite entanglement $Eom$, $EoM$ and $EmM$. We remark that the entanglement in the Fig. \ref{dynamics}(a) exhibits three regimes of the entanglement of the all three entanglement. The first regime concerns classically correlated states (zero entanglement), i.e. when two are separated. This indicates the absence of any quantum correlations transfer between two modes. The second regime corresponds to the emergence of entanglement between the two modes. In this regime we observe the generation of the oscillation in time this can be explained by the Sørensen-Mølmer entanglement dynamics discussed in Refs. \cite{JLi2017}. The third regime, corresponds to large periods of evolution and associated with the entanglement between the two modes when they reach the steady regime. We remark in Fig. \ref{dynamics}(b) that the system is a genuinely tripartite entangled state as shown by the nonzero minimum residual contangle $R_{min}$.

\section{Conclusions} \label{Conc}

In summary, we have proposed a theoretical scheme to enhance the three bipartite and tripartite entanglement in optomagnomechanical system. We have quantified the amount of entanglement in all bipartite and tripartite mode via logarithmic negativity. We have shown the genuinely tripartite entanglement state via the nonzero minimum residual contangle $\mathcal{R}_{min}$. Besides, we have discussed the behavior of stationary and dynamics of three bipartite and tripartite entanglement versus the parameter reflective of beam splitter and the phenomenon of decoherence effects using experimentally feasible parameters. We have shown that the presence of coherent feedback loop enhances the bipartite photon-magnon and magnon-phonon entanglement $Eom$ and $EmM$, respectively. Moreover, the presence of coherent feedback loop degrade the photon-phonon entanglement $EoM$ as shown in Fig. \ref{bimodes}(c). Our results show that the entanglement is fragile under thermal (decoherence) effects while the robustness of entanglement in the presence of coherent feedback can be achieved.

\end{document}